\documentstyle[psfig]{natloc}
\title{The two-hour orbit of a binary millisecond X-ray pulsar}

\author{Deepto Chakrabarty \& Edward H. Morgan 
\institute{Center for Space Research, Massachusetts Institute of
Technology, Cambridge, MA 02139 USA}} 

\headertitle{}
\mainauthor{}

\summary{Typical radio pulsars are magnetized neutron stars that are
born rapidly rotating and slow down as they age on time scales of 10
to 100 million years.  However, millisecond radio pulsars spin very
rapidly even though many are billions of years old\cite{bkh+82}. 
The most compelling explanation is that they have been ``spun up'' by
the transfer of angular momentum during accretion of material from a
companion star in so-called low-mass X-ray binary systems, LMXBs. 
(LMXBs consist of a neutron star or black hole accreting from a
companion less than one solar mass\cite{bv91}.)  The recent detection
of coherent X-ray pulsations with a millisecond period from a
suspected low-mass X-ray binary system appears to confirm this
link\cite{wv98b}.  Here we report observations showing that the
orbital period of this binary system is two hours, which establishes
it as an LMXB.  We also find an apparent modulation of the X-ray
flux at the orbital period (at the two per cent level), with a 
broad minimum when the pulsar is behind this low-mass companion star.
This system seems closely related to the ``black widow'' millisecond
radio pulsars, which are evaporating their companions through 
irradiation\cite{fbb+90,nt92,mlr+91,dma+93,sbl+96}.  It may appear 
as an eclipsing radio pulsar during periods of X-ray quiescence. }

\dates{23 April 1998}{16 June 1998}

\begin{document}
\maketitle

The transient X-ray source SAX~J1808.4--3658 was first observed in
September 1996 by the Wide Field Cameras on the BeppoSAX X-ray
satellite during a bright state which lasted about 20
days\cite{ihm+98}.   The source was not detected (X-ray flux $<3\times
10^{-11}$ erg cm$^{-2}$ s$^{-1}$ in the 2--10 keV band) during an August 1996
observation, but reached a peak X-ray flux of $2\times 10^{-9}$ erg
cm$^{-2}$ s$^{-1}$ during the September bright state.  Also during the
bright state, BeppoSAX detected two type I X-ray bursts from the
source, each lasting less than 30 s.  Such bursts are due to
thermonuclear ignition of accreted material on the surface of neutron
stars that have low ($<10^{10}$ G) magnetic fields \cite{lvt95}.  Analysis
of the bursts in SAX~J1808.4--3658 indicates that it is 4 kpc distant
and has a peak X-ray luminosity of $6\times 10^{36}$ erg~s$^{-1}$ in
its bright state and $<10^{35}$ erg s$^{-1}$ in
quiescence\cite{ihm+98}.  The source position was determined to within
a 1$\sigma$ radius of 0.6 arcmin.

More recently, a serendipitous slew of the Proportional Counter Array
(PCA) on the Rossi X-Ray Timing Explorer (RXTE) over this region of the sky
on 9 April 1998 indicated the presence of an X-ray source, designated
XTE~J1808--369.  Repeated scans over the region with the PCA on 11
April localized the position with sufficient precision to confirm that
the source was probably the same as SAX~J1808.4--3658\cite{mar98}.
RXTE subsequently made numerous pointed observations of the source
during 11 April--6 May on a public ``target-of-opportunity'' basis.
The source reached a peak luminosity of $5\times 10^{36}$ erg s$^{-1}$
on 13 April and had faded below $10^{35}$ erg s$^{-1}$ by 6 May.  

Strong coherent 400 Hz pulsations were clearly detected in the 2--30
keV PCA data from most of these observations, and were first reported
by Wijnands and van der Klis\cite{wv98a,wv98b}.  These are the fastest
persistent coherent pulsations ever detected from an X-ray binary.
From standard magnetic disk accretion theory, the detection of X-ray
pulsations at a luminosity $\sim 10^{36}$ erg s$^{-1}$ requires that
the neutron star's surface dipole magnetic field be $B< 10^8$ G
(ref. \pcite{wv98b}).  This is much weaker than the $\sim 10^{12}$ G
fields found in other (slower) accretion-powered pulsars\cite{wnp95},
but it is consistent with the weak fields expected for type~I X-ray
bursters\cite{lvt95}.  If pulsations are detected when the source is
at lower luminosities, the implied surface field would be even weaker.

In order to search for orbital Doppler shifts, we first
selected the 3--30 keV photon arrival times during 11--18 April and
binned them into $2^{-11}$ s ($\approx$0.5 ms) samples, after
correcting these times for RXTE's motion with respect to the solar
system barycenter using the best BeppoSAX position for the
source\cite{ihm+98} (right ascension 18h 08m 29s, declination
$-36^\circ$ 58.6$'$, equinox J2000.0).   We then measured the
barycentric pulse frequency at 128~s intervals using an oversampled
Fourier power spectrum.  The sequence of pulse frequencies showed an
obvious 2 hr sinusoidal modulation which was well fit by a constant
spin frequency plus a circular, Keplerian orbit\cite{cm98}.   

Using this provisional timing model, we epoch-folded each 128 s
interval in the 11--18 April data and cross-correlated the resulting
pulse profiles with a sinusoid to measure the pulse phase history,
which generally yields more precise timing parameters than a pulse
frequency history.  These pulse phases were well fit (reduced
$\chi^2$=1.26 with 441 degrees of freedom) by a constant spin
frequency $\nu$ plus a circular Keplerian orbit.  The best-fit
parameters are given in Table~1.  The pulse arrival time delays with
respect to a constant frequency model are plotted in Figure~1.  No
improvement in the fit to the 11--18 April data is obtained by
including an orbital eccentricity $e$ or a spin frequency derivative
$\dot\nu$.   A more sensitive search for a non-zero value of
$\dot\nu$, which includes the later data taken as the X-ray outburst
faded from view, will be presented elsewhere. 

The present limits on $\dot\nu$ constrain the positional uncertainty
of the star.  An error of 0.6 arcmin in the assumed source position
would give rise to periodic timing residuals with 1~yr period and
$\sim$80 ms amplitude\cite{lg90}, and such residuals would lead
to an apparent $\dot\nu$ in the timing solution for observations
shorter than 3 months.  Our limit on $\dot\nu$ from the 11--18 April
data suggests that the BeppoSAX position is accurate to within 25
arcsec in ecliptic longitude.  We note that a probable optical
counterpart has been identified, lying 19 arcsec from the BeppoSAX
position\cite{rcm+98,ghg98}. 

\begin{table}
\caption{Observed parameters of SAX J1808.4--3658}
{\small
\begin{tabular}{ll}\hline
Barycentric pulse frequency, $\nu_0$ (Hz)             & 400.9752106(8) \\
Pulse frequency derivative,  $|\dot\nu|$ (Hz s$^{-1}$)& $< 7\times 10^{-13}$ \\
Projected semimajor axis, $a_1\sin i$ (lt-ms)         & 62.809(1)\\
Orbital period, $P_{\rm orb}$ (s)                        & 7249.119(1) \\
Epoch of 90$^\circ$ mean longitude, $T_{\pi/2}$ (MJD) & 50914.899440(1) \\
Eccentricity, $e$ & $<5\times 10^{-4}$ \\
Pulsar mass function, $f_1$ ($M_\odot$)	          & $3.7789(2)\times 10^{-5}$\\
\hline
\end{tabular}

Modified Julian date MJD = JD $-$ 2400000.5.  Numbers in parentheses are
the 1$\sigma$ uncertainties in the last significant figure, and upper
limits are quoted at the $2\sigma$ level. The $\dot\nu$ limit refers
to the magnitude, independent of sign.  Epochs are given in Barycentric
Dynamical Time (TDB).  
}
\end{table}

\begin{figure}[t]
\psfig{file=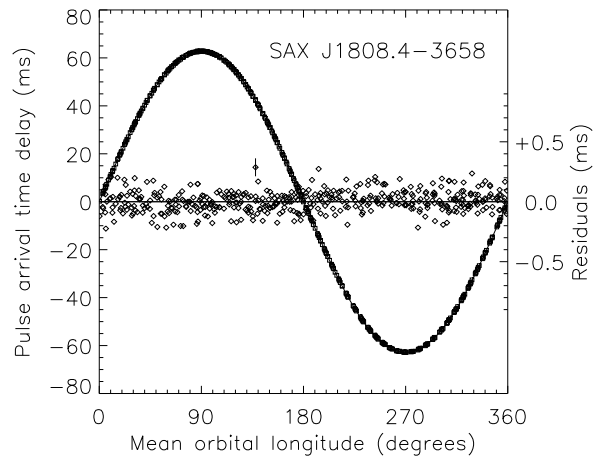,width=3in}
\caption{Pulse arrival time delays due to the 2~hr binary orbit, with
respect to a constant pulse frequency model.  Mean longitude is the
orbital phase angle measured from the ascending node, so that the
pulsar is behind the companion at 90$^\circ$.  The squares are the
measured time delays, the solid curve is the best-fit orbital model,
and the diamonds are the model residuals (on a 50$\times$ expanded
scale).  The residuals have an r.m.s. uncertainty of 75 $\mu$s, shown
for one of the points.} 
\end{figure}

The pulsar mass function $f_1$, which relates the pulsar mass $m_1$,
the companion mass $m_2$, and the binary inclination $i$ (the angle
between the line of sight and the orbital angular momentum vector),
may be computed from the observed Keplerian parameters $a_1\sin i$ and
$P_{\rm orb}$, 
\begin{equation}
f_1\equiv \frac{(m_2\sin i)^3}{(m_1+m_2)^2}
   = \frac{4\pi^2 (a_1\sin i)^3}{G P_{\rm orb}^2} ,
\end{equation}
from which we find $f_1= 3.7789(2)\times 10^{-5}\,M_\odot$ (here
$M_\odot$ is the solar mass).  Given assumed
values of $m_1$ and $i$, equation~(1) can be solved for $m_2$.  For
$m_1$, dynamical mass measurements in both binary millisecond radio
pulsars\cite{tc98} and accretion-powered pulsars\cite{vvz95} are all
generally consistent with a neutron star mass of 1.35 $M_\odot$.
However, neutron stars in LMXBs
have undergone substantial accretion from their binary
companion and may be as massive as 2 $M_\odot$ (e.g.,
ref. \pcite{zss97}).  For $i$, we note that for an ensemble of
binaries whose inclinations are distributed randomly, the 
{\em a priori} probability of observing a system with inclination $i$
or smaller is $(1-\cos i)$.   If SAX J1808.4--3658 is drawn from such
an ensemble, then there is a 95\% probability that $m_2<0.14\,M_\odot$
if $m_1=1.35\,M_\odot$.   For $m_1=2\,M_\odot$, we find
$m_2<0.18\,M_\odot$.  In either case, the binary separation is
most likely of order 1~lt-s. 

The nature of the companion star provides an important clue to the
evolutionary history of an LMXB.  The only direct information we have
on the companion comes from observations of the probable optical
counterpart, which suggest that the companion is a faint low-mass star
subject to X-ray heating by the pulsar\cite{rcm+98,ghg98}.  We can use
the binary parameters to deduce more about this star.  The companions
in most LMXBs fill (or nearly fill) their critical gravitational
potential surface, known as the Roche lobe.  Because the mean density
of a Roche-lobe--filling companion (for the case where $m_2<m_1$) is
uniquely determined by the binary period\cite{ffw72}, we can relate
the companion's radius in SAX J1808.8--3658 to its mass as
$R_2 = 0.17\,R_\odot\,(m_2/0.1\, M_\odot)^{1/3}$, giving a minimum companion
radius of 0.12 $R_\odot$ (here $R_\odot$ is the solar radius).  The
assumption of Roche lobe overflow thus 
strongly constrains the nature of the companion.  Given the compact
size of the binary, only white dwarf (WD) or low-mass main sequence
(MS) companions are plausible.  (In principle, a helium-burning star
could also fit in this binary, but we would expect it to be much
brighter than the proposed optical counterpart\cite{sdv86}.)  A WD
with mass $m$ has radius $R=0.013\, R_\odot (1+X)^{5/3}
(m/M_\odot)^{-1/3}$, where $X$ is the hydrogen mass
fraction\cite{pac67}.  Comparing this to our equation for $R_2$, we see
that a helium WD ($X=0$) would be too small, and hence is ruled out.
Even a hydrogen-rich WD ($X=0.9$) is essentially excluded for all but
$i\approx 90^\circ$. However, the lack of a deep X-ray eclipse rules
out $i>82^\circ$, assuming a Roche-lobe--filling companion. 

Normal low-mass hydrogen MS stars have $R/R_\odot\approx m/M_\odot$,
and comparison of detailed models\cite{tpe+96} with our equation for $R_2$
yields a Roche-lobe--filling mass of 0.17 $M_\odot$.
This would require a small binary inclination ($i<20^\circ$), which
has low {\em a priori} probability (5\%).  However, normal MS 
models are inappropriate in our case, since the stellar structure must
be strongly influenced by irradiation from the pulsar.  If the
incident flux at the companion surface exceeds $\sim 10^{10}$
erg~cm$^{-2}$~s$^{-1}$, the star may be ``bloated'' to larger
radii\cite{pod91}.  This is especially true for MS stars having
$m<0.3\, M_\odot$, which have fully convective envelopes.  For the
$\sim$1 lt-s binary separation in this system, the X-ray flux at the
companion will exceed $10^{14}$ erg~cm$^{-2}$~s$^{-1}$, and the
bloating can be severe.  Under these conditions, MS stars as light as
0.1\,$M_\odot$ or less may fill their Roche lobe, consistent with a more
likely inclination.

Another possibility is that irradiation by the neutron star might
directly drive mass loss from the companion (``ablation'') whether or
not it fills its Roche lobe\cite{dan96}.  The binary parameters of
SAX J1808.4--3658 are very similar to those of the five ``black
widow'' millisecond radio pulsars in very close binaries, all of which are 
ablating their low mass companions\cite{fbb+90,nt92,mlr+91,dma+93,sbl+96}.
At least four of these radio systems show eclipses which are too broad to 
be caused by a Roche-lobe--filling companion, and which are instead
attributed to an ablated wind\cite{fbb+90,nt92,mlr+91,sbl+96}.

We find evidence that the X-ray flux from SAX
J1808.4--3658 is slightly modulated at the orbital period.  The apparent
modulation is roughly sinusoidal with 2\% amplitude and a minimum
when the pulsar is behind the companion, and it is much broader and
much shallower than the (at most) 5 minute eclipse possible from a
Roche-lobe--filling companion.   However, the detailed features of
this 2~hr modulation must be viewed with caution, since its strength
is comparable to the variation of the RXTE background with the 96 min
spacecraft orbit.  The similarity in strength and time scales of these
two variations makes them difficult to disentangle, and final
confirmation must await either additional data or an improvement in 
the RXTE background model.  Still, the detection of a 2~hr source flux
modulation (if not its precise morphology) seems secure.  

We suggest that the intensity dips are due to scattering in an ablated
wind.  Given the similarity between this source and the eclipsing
radio pulsars, SAX J1808.4--3658 may emerge as a radio pulsar during
X-ray quiescence\cite{scc+94}.  Moreover, since the radio emission would
be less penetrating than the X-rays, a deeper eclipse dip might be
observed, possibly providing a strong constraint on the binary
inclination.  The presence of the slight X-ray dips, if confirmed,
would rule out small binary inclinations and make it highly likely
that the companion mass is less than 0.1 $M_\odot$.

Whether the mass accretion is fed by Roche-lobe overflow or an ablated
wind or both, we can understand the transient nature of the X-ray
emission as long as some of this material forms an accretion disk.
Although the instantaneous mass accretion rate during the 1996 and
1998 bright states was $\approx 3\times 10^{-10}\,M_\odot$~yr$^{-1}$,
the small duty cycle of the X-ray activity indicates a long-term mean
mass transfer rate of $\dot M\approx 1\times 10^{-11}\,M_\odot$~yr$^{-1}$.
For such a low $\dot M$, the accretion disk would probably be subject
to dwarf-nova--type instabilities, leading to episodic outbursts of
X-ray emission\cite{van96,kkb96}.  We note that for a
Roche-lobe--filling companion, angular momentum losses due to
gravitational radiation would drive a mass transfer rate of $\approx
10^{-11}\, M_\odot$ yr$^{-1}$ for a 0.05 $M_\odot$
secondary\cite{vv95}, consistent with our inferred inclination
constraints.

Future X-ray outbursts from SAX J1808.4--3658 may allow detection of a
spin frequency derivative, which would further constrain both the mass
accretion rate and the pulsar's magnetic field strength and would provide
a probe of magnetic disk accretion torque theory in the previously
unexplored regime of a very small magnetosphere.  The characteristic
accretion torque expected for the observed X-ray luminosity should cause
a $\dot\nu$ of order $10^{-14}$ Hz~s$^{-1}$, which is detectable in 
observations spanning a month or more.  Further observations may also
detect orbital period evolution, which would probe the competing
effects of gravitational radiation, tidal interactions, and mass loss
on the binary angular momentum.  An astrophysically interesting limit
of $|\dot P_{\rm orb}| < 10^{-11}$ could be reached in less than a
year of observations.

\bibliographystyle{nature}

\smallskip
\noindent {\small {\bf Acknowledgements.} We thank L. Bildsten,
V. Kaspi, A. Levine, R. Nelson, R. Remillard, F. Rasio, S.
Thorsett, M. van der Klis, and B. Vaughan for useful
discussions, and M. Muno for assistance with the data analysis.  We
also thank F. Marshall, J. Swank and the RXTE team at
NASA/Goddard Space Flight Center for arranging these
target-of-opportunity observations and the necessary follow-up.  This
work was supported by NASA.}

\medskip
\noindent {\small Correspondence should be addressed to D.C. (e-mail:
deepto@space.mit.edu).} 

\end{document}